\documentclass[pra,tightenlines,superscriptaddress,12pt]{revtex4}

\newtheorem{definition}{Definition}
\newtheorem{theorem}{Theorem}

\begin{document}

\def\ket#1{|#1\rangle}
\def\bra#1{\langle#1|}
\def\tr{{\rm tr}}

\def\HD{{\cal H}_D}
\def\H{{\cal H}}

\title{A de Finetti Representation Theorem for Quantum Process Tomography}

\author{Christopher A.~Fuchs}
\email{cafuchs@research.bell-labs.com}
\affiliation{Bell Labs, Lucent Technologies, 600--700 Mountain Avenue,
Murray Hill, NJ 07974, USA}
\affiliation{Communication Networks Research Institute,
Dublin Institute of Technology, Rathmines Road, Dublin 6, Ireland}
\affiliation{Department of Physics and Special Research Centre for
Quantum
  Computer Technology, The University of Queensland, Queensland 4072,
  Australia}
\author{R\"udiger Schack}
\email{r.schack@rhul.ac.uk}
\affiliation{Department of Mathematics, Royal Holloway, University of
London, Egham, Surrey TW20$\;$0EX, UK}
\affiliation{Department of Physics and Special Research Centre for Quantum
  Computer Technology, The University of Queensland, Queensland 4072,
  Australia}
\author{Petra F. Scudo}
\email{pscudo@libero.it}
\affiliation{Bell Labs, Lucent
Technologies, 600--700 Mountain Avenue, Murray Hill, NJ 07974,
USA} \affiliation{Department of Physics, Technion---Israel
Institute of Technology, 32000 Haifa, Israel}
\date{19 July 2003}

\begin{abstract}
\vspace{.1in}
In quantum process tomography, it is possible to express the
experimenter's prior information as a sequence of quantum operations,
i.e., trace-preserving completely positive maps. In analogy to de
Finetti's concept of exchangeability for probability distributions,
we give a definition of exchangeability for sequences of quantum
operations. We then state and prove a representation theorem for such
exchangeable sequences. The theorem leads to a simple
characterization of admissible priors for quantum process tomography
and solves to a Bayesian's satisfaction the problem of an {\it
unknown\/} quantum operation.
\end{abstract}


\maketitle

\section{Introduction}

In quantum process
tomography~\cite{Turchette1995b,Chuang1997,Poyatos1997}, an
experimenter lets an incompletely specified device act on a quantum
system prepared in an input state of his choice, and then performs a
measurement (also of his choice) on the output system. This procedure
is repeated many times over, with possibly different input states and
different measurements, in order to accumulate enough statistics to
assign a quantum operation to the device. Here and throughout the
paper, by a quantum operation we mean a trace-preserving completely
positive linear map---the most general description for
(unconditioned) quantum-state evolution allowed by the laws of
quantum mechanics~\cite{Nielsen2000}. Quantum process tomography has
been demonstrated experimentally in liquid state nuclear magnetic
resonance \cite{Nielsen1998a,Childs2001}, and recently a number of
optical experiments \cite{Nambu2002,DeMartini-0210,Altepeter-0303}
have implemented entanglement-assisted quantum process tomography.
The latter is a procedure that exploits the fact that quantum process
tomography is equivalent to quantum state tomography in a larger
state space \cite{Leung2001,Leung2002,Duer2001,D'Ariano2001}.

In the usual description of process tomography, it is assumed that
the device performs the same {\it unknown\/} quantum operation $\Phi$
every time it is used, and an experimenter's prior information about
the device is expressed via a probability density $p(\Phi)$ over all
possible operations.  What, however, is the operational meaning of an
unknown quantum operation?  When does the action of a device leave
off from an initial input so that the next input can be sent through?
In particular, what gives the right to suppose that a device does not
have memory or, for instance, does not entangle the successive inputs
passing through it?  These questions boil down to the need to explore
a single issue: What essential assumptions must be made so that
quantum process tomography is a logically coherent notion?

In this paper, we address this issue with a uniqueness theorem based
on (quantum) Bayesian methodology
\cite{Schack2001a,Caves2002a,Caves2002b,Fuchs2002,Caves2002c,Brun2001}.
What is called for is a method of posing quantum process tomography
that never requires the invocation of the concept of an unknown
quantum operation. This can be done by focussing upon the action of a
single {\it known\/} quantum operation $\Phi^{(N)}$, which acts upon
$N$ nominal inputs.  In particular, we identify conditions under
which $\Phi^{(N)}$, ($N=1,2,\ldots$), can be represented as
\begin{equation}   \label{eq:prior}
\Phi^{(N)}=\int d\Phi\; p(\Phi)\, \Phi^{\otimes N} \;,
\end{equation}
for some probability density $p(\Phi)$, and where the integration
extends over all single-system quantum operations $\Phi$.  With this
theorem established, the conditions under which an experimenter can
act {\it as if\/} his prior $\Phi^{(N)}$ corresponds to {\em
ignorance\/} of a ``true'' but unknown quantum operation are made
precise.

Our starting point is the closely aligned and similarly motivated de
Finetti representation theorem for quantum states
\cite{Hudson1976,Caves2002b,Emch2002}.   According to this theorem, a
state $\rho^{(N)}$ of $N$ systems can be written in the form
\begin{equation}   \label{eq:definetti}
\rho^{(N)}=\int d\rho\; p(\rho)\, \rho^{\otimes N}
\label{BigBoy}
\end{equation}
if and only if $\rho^{(N)}$ is an element of an {\it exchangeable
sequence}. A quantum state $\rho^{(N)}$ of $N$ systems is said to be
a member of an exchangeable sequence if
\begin{verse}
(i) $\rho^{(k)}$ is symmetric, i.e., is invariant under permutations
of the $k$ systems on which it is defined, and\medskip \\
(ii) $\rho^{(k)}=\tr_{k+1}\rho^{(k+1)}$, where $\tr_{k+1}$ denotes
the partial trace over the $(k+1)$th system.
\end{verse}
In representation (\ref{BigBoy}), $d\rho$ is a suitable measure on
the density operator space, and $p(\rho)\ge0$ is unique. The concept
of exchangeability \cite{Regazzini1988} was first introduced by Bruno
de Finetti for sequences of probability distributions.

Here, we make use of the correspondence between quantum process
tomography and quantum state tomography mentioned above to derive a
de Finetti representation theorem for sequences of quantum
operations. In Sec.~\ref{sectheorem} we define exchangeability for
quantum operations and state the theorem. The proof is given in
Sec.~\ref{secproof}.  We close the paper with some concluding remarks
that emphasize the quantum foundational character of our result.

\section{The theorem}
\label{sectheorem}

In this paper, we restrict our attention to devices for which the
input and output have the same Hilbert space dimension, $D$.  In the
following, $\HD$ denotes a $D$-dimensional Hilbert space,
$\HD^{\otimes N}=\HD\otimes\cdots\otimes\HD$ denotes its $N$-fold
tensor product, and ${\cal L}({\cal V})$ denotes the space of linear
operators on a linear space ${\cal V}$. The set of density operators
for a $D$-dimensional quantum system is a convex subset of ${\cal
  L}(\HD)$.

The action of a device on $N$ nominal inputs systems is then
described by a trace-preserving completely positive map
\begin{equation}
\Phi^{(N)} : {\cal L}(\HD^{\otimes N}) \longrightarrow
  {\cal L}(\HD^{\otimes N}) \;,
\end{equation}
which maps the state of the $N$ input systems to the state of the $N$ output
systems. We will say, in analogy to the definition of exchangeability for
quantum states, that a quantum operation $\Phi^{(N)}$ is {\it
  exchangeable\/} if it is a member of an exchangeable sequence of
quantum operations.

To define exchangeability for a sequence of quantum operations in a natural
way, we reduce the properties of symmetry and extendibility for sequences of
operations to the corresponding properties for sequences of states. In the
following, we will use bold letters to denote vectors of indices, e.g.  ${\bf
  j}=(j_1,\ldots,j_N)$. We will use $\pi$ to denote a permutation of the set
$\{1,\ldots,N\}$, where the cardinality $N$ will depend on the
context. The action of the permutation $\pi$ on the vector ${\bf j}$
is defined by $\pi{\bf j} = (j_{\pi(1)},\ldots,j_{\pi(N)})$.

Any $N$-system density operator $\rho^{(N)}$ can  be expanded in the
form
\begin{equation}
\rho^{(N)}=\sum_{{\bf j},{\bf l}} r^{(N)}_{{\bf j},{\bf l}}
\bigotimes_{i=1}^N |j_i^{Q_i}\rangle\langle l_i^{Q_i}| \equiv
\sum_{{\bf j},{\bf l}} r^{(N)}_{{\bf j},{\bf l}}
|j_1^{Q_1}\rangle\langle l_1^{Q_1}|
  \otimes\cdots\otimes |j_N^{Q_N}\rangle\langle l_N^{Q_N}|  \;,
\end{equation}
where $\{\ket{1^{Q_i}},\ldots,\ket{D^{Q_i}}\}$ denotes an
orthonormal basis for the Hilbert space $\HD$ of the $i$th system,
and $r^{(N)}_{{\bf j},{\bf l}}$ are the matrix elements of
$\rho^{(N)}$ in the tensor product basis. We define the action of
the permutation $\pi$ on the state $\rho^{(N)}$ by
\begin{equation}
\pi\rho^{(N)}=\sum_{{\bf j},{\bf l}} r^{(N)}_{\pi{\bf j},\pi{\bf l}}
\bigotimes_{i=1}^N |j_i^{Q_i}\rangle\langle l_i^{Q_i}|
=\sum_{{\bf j},{\bf l}} r^{(N)}_{{\bf j},{\bf l}}
\bigotimes_{i=1}^N |j_{\pi^{-1}(i)}^{Q_i}\rangle\langle l_{\pi^{-1}(i)}^{Q_i}|
\;.
\end{equation}
With this notation, we can make the following definition.

\begin{quote}
\begin{definition}
A sequence of quantum operations, $\Phi^{(k)} : {\cal L}(\HD^{\otimes
k}) \rightarrow {\cal L}(\HD^{\otimes k})$, is exchangeable if, for
$k=1,2,\ldots$,

\noindent(i) $\Phi^{(k)}$ is symmetric, i.e.,
\begin{equation}   \label{eq:symmetric}
\Phi^{(k)}(\rho^{(k)}) = \pi\Big(\Phi^{(k)}(\pi^{-1}\rho^{(k)})\Big)
\end{equation}
for any permutation $\pi$ of the set $\{1,\ldots,k\}$ and for any density
operator $\rho^{(k)} \in {\cal L}(\HD^{\otimes k})$,
and

\noindent(ii) $\Phi^{(k)}$ is extendible, i.e.,
\begin{equation}  \label{eq:extendible}
\Phi^{(k)}(\tr_{k+1}\rho^{(k+1)}) =
\tr_{k+1}\Big(\Phi^{(k+1)}(\rho^{(k+1)})\Big)
\end{equation}
for any state $\rho^{(k+1)}$.
\end{definition}
\end{quote}

In words, these conditions amount to the following. Condition (i) is
equivalent to the requirement that the quantum operation $\Phi^{(k)}$
commutes with any permutation operator $\pi$ acting on the states
$\rho^{(k)}$:  It does not matter what order we send our systems
through the device; as long as we rearrange them at the end into the
original order, the resulting evolution will be the same. Condition
(ii) says that it does not matter if we consider a larger map
$\Phi^{(N+1)}$ acting on a larger collection of systems (possibly
entangled), or a smaller $\Phi^{(N)}$ on some subset of those
systems:  The upshot of the evolution will be the same for the
relevant systems.

We are now in a position to formulate the de Finetti
representation theorem for quantum operations.

\begin{quote}
\begin{theorem}
A quantum operation $\Phi^{(N)} : {\cal L}(\HD^{\otimes N})
\rightarrow {\cal L}(\HD^{\otimes N})$ is an element of an
exchangeable sequence if and only if it can be written in the form
\begin{equation}
\Phi^{(N)}=\int d\Phi\; p(\Phi)\, \Phi^{\otimes N} \, \quad \mbox{for
all $N$},
\end{equation}
where the integral ranges over all single-shot quantum operations
$\Phi:{\cal L}(\HD)\rightarrow{\cal L}(\HD)$, $d\Phi$ is a suitable
measure on the space of quantum operations, and the probability
density $p(\Phi)\ge0$ is unique. The tensor product $\Phi^{\otimes
N}$ is defined by $\Phi^{\otimes N}(\rho_1\otimes\cdots\otimes\rho_N)
=\Phi(\rho_1)\otimes\cdots\otimes\Phi(\rho_N)$ for all
$\rho_1,\ldots,\rho_N$ and by linear extension for arbitrary
arguments.
\end{theorem}
\end{quote}

Just as with the original quantum de Finetti theorem
\cite{Hudson1976,Caves2002b}, this result allows a certain latitude
in how quantum process tomography can be described.  One is free to
use the language of an unknown quantum operation if the condition of
exchangeability is met by one's prior $\Phi^{(N)}$ but it is not
required:  For the (quantum) Bayesian statistician the {\it known\/}
quantum operation $\Phi^{(N)}$ is the more fundamental object.

\section{Proof}  \label{secproof}

Let $\Phi^{(N)}$, $N=1,2,\ldots$, be an exchangeable sequence of
quantum operations. $\Phi^{(N)}$ can be
characterized in terms of its action on the elements of a basis of
${\cal L}(\HD^{\otimes N})$ as follows.
\begin{equation}   \label{eq:sljmk}
\Phi^{(N)} \Big( \bigotimes_{i=1}^N \ket{j_i^{Q_i}}\bra{k_i^{Q_i}} \Big)
=\sum_{{\bf l},{\bf m}}
 S^{(N)}_{{\bf l},{\bf j},{\bf m},{\bf k}}
       \bigotimes_{i=1}^N \ket{l_i^{Q_i}}\bra{m_i^{Q_i}} \;.
\end{equation}
The coefficients
$S^{(N)}_{{\bf l},{\bf j},{\bf m},{\bf k}}$
specify $\Phi^{(N)}$ uniquely.  It follows from a construction due to Jamio\l
kowski \cite{Jamiolkowski1972} that the $S^{(N)}_{{\bf l},{\bf j},{\bf m},{\bf
    k}}$ can be regarded as the matrix elements of a density operator on
$D^{2N}$-dimensional Hilbert space $\H_{D^2}^{\otimes N}$. This can be seen as
follows.  Let
\begin{equation}
\ket\Psi = {1\over\sqrt D} \sum_{k=1}^D \ket{k^{R_i}}\ket{k^{Q_i}}
   \in \HD\otimes\HD= \H_{D^2}
\end{equation}
be a maximally entangled state in $\H_{D^2}$, where the
$\ket{k^{R_i}}$ ($k=1,\ldots,D$) form orthonormal bases for the
ancillary systems labelled $R_i$ ($i=1,\ldots,N$). The corresponding
density operator is
\begin{equation}
\ket\Psi\bra\Psi =
{1\over D} \sum_{j,k} \ket{j^{R_i}}\bra{k^{R_i}}
   \otimes\ket{j^{Q_i}}\bra{k^{Q_i}} \in {\cal
  L}(\H_{D^2}) \;.
\end{equation}
Similarly, we define a map, $J$, from the set
of quantum operations on $\HD^{\otimes N}$ to the set of
density operators on $\H_{D^2}^{\otimes N}$ by
\begin{eqnarray}   \label{eq:Jdef}
J ( \Phi^{(N)} ) &\equiv&
\Big(I^{(N)} \otimes \Phi^{(N)}\Big)
   \Big( (\ket\Psi\bra\Psi)^{\otimes N} \Big)  \cr
&=&
{1\over D^N}\Big(I^{(N)} \otimes \Phi^{(N)}\Big)
    \Big( \sum_{{\bf j},{\bf k}} \bigotimes_{i=1}^N
(\ket{j_i^{R_i}}\bra{k_i^{R_i}} \otimes \ket{j_i^{Q_i}}\bra{k_i^{Q_i}}) \Big)
  \cr
&=&
{1\over D^N} \sum_{{\bf l},{\bf j},{\bf m},{\bf k}}
 S^{(N)}_{{\bf l},{\bf j},{\bf m},{\bf k}} \bigotimes_{i=1}^N
(\ket{j_i^{R_i}}\bra{k_i^{R_i}} \otimes \ket{l_i^{Q_i}}\bra{m_i^{Q_i}}) \;.
\end{eqnarray}
In this definition, $I^{(N)}$ denotes the identity operation acting on the
ancillary systems $R_1,\ldots,R_N$.  The map $J$ is injective, i.e.
$J(\Phi_1^{(N)})=J(\Phi_2^{(N)})$ if and only if $\Phi_1^{(N)}=\Phi_2^{(N)}$.

The first stage of the proof of the de Finetti theorem for operations is to
show that the density operators $J(\Phi^{(N)})$, $N=1,2,\ldots$, form an
exchangeable sequence when regarded as $N$-system states, with $R_i$ and $Q_i$
jointly forming the $i$th system.  To do this, we first show that
$J(\Phi^{(N)})$ is symmetric, i.e., invariant under an arbitrary permutation
$\pi$ of the $N$ systems.

Note that since the density operators $\rho^{(N)}$ actually span the
whole vector space ${\cal L}(\HD^{\otimes N})$, enforcing Definition
1 above amounts to identifying the linear maps on the left- and
right-hand sides of Eqs.~(\ref{eq:symmetric}) and
(\ref{eq:extendible}).  I.e.,
\begin{equation}
\Phi^{(k)} = \pi\circ\Phi^{(k)}\circ\pi^{-1}
\end{equation}
and
\begin{equation}
\Phi^{(k)}\circ\tr_{k+1} = \tr_{k+1}\circ\Phi^{(k+1)}
\end{equation}
Thus in much that we do it suffices to consider the action of these
maps on an arbitrary basis state $E^{(N)}=\bigotimes_{i=1}^N
\ket{j_i^{Q_i}}\bra{k_i^{Q_i}}$ for arbitrary ${\bf j}$ and ${\bf
k}$. In particular,
\begin{eqnarray}   \label{eq:pisljmk}
\pi\Big(\Phi^{(N)}(\pi^{-1}E^{(N)})\Big)
&=&
\pi\Big(\Phi^{(N)}\Big(
 \bigotimes_{i=1}^N \ket{j_{\pi(i)}^{Q_i}}\bra{k_{\pi(i)}^{Q_i}}
 \Big)\Big)  \cr
&=&
\pi\sum_{{\bf l},{\bf m}}
 S^{(N)}_{{\bf l},\pi{\bf j},{\bf m},\pi{\bf k}}
       \bigotimes_{i=1}^N \ket{l_i^{Q_i}}\bra{m_i^{Q_i}} \cr
&=&
\sum_{{\bf l},{\bf m}}
 S^{(N)}_{\pi{\bf l},\pi{\bf j},\pi{\bf m},\pi{\bf k}}
       \bigotimes_{i=1}^N \ket{l_i^{Q_i}}\bra{m_i^{Q_i}} \;.
\end{eqnarray}
Assuming Eq.~(\ref{eq:symmetric}), i.e.,
symmetry of $\Phi^{(N)}$, for all ${\bf j}$ and ${\bf k}$, it follows that
\begin{equation}
S^{(N)}_{\pi{\bf l},\pi{\bf j},\pi{\bf m},\pi{\bf k}}
=  S^{(N)}_{{\bf l},{\bf j},{\bf m},{\bf k}}
\end{equation}
for all ${\bf l},{\bf j},{\bf m},{\bf k}$, which, using
Eq.~(\ref{eq:Jdef}), implies that
\begin{equation}
\pi(J ( \Phi^{(N)} )) =  J ( \Phi^{(N)} ) \;,
\end{equation}
i.e., symmetry of $J(\Phi^{(N)})$.

To prove extendibility of $J(\Phi^{(N)})$, we introduce the
following notation for partial traces: we denote by $\tr_{N+1}^R$
the partial trace over the subsystem $R_{N+1}$, and by
$\tr_{N+1}^Q$ the partial trace over the subsystem $Q_{N+1}$. In
this notation, we need to show that $\tr_{N+1}^R\tr_{N+1}^Q
J(\Phi^{(N+1)})=J(\Phi^{(N)})$. Using Eqs.~(\ref{eq:extendible})
and~(\ref{eq:Jdef}),
\begin{eqnarray}
&& \tr_{N+1}^R\tr_{N+1}^Q J(\Phi^{(N+1)}) \cr
&& =
\tr_{N+1}^R\tr_{N+1}^Q{1\over D^{N+1}}\Big(I^{(N+1)} \otimes \Phi^{(N+1)}\Big)
    \Big( \sum_{{\bf j},j_{N+1},{\bf k},k_{N+1}} \bigotimes_{i=1}^{N+1}
(\ket{j_i^{R_i}}\bra{k_i^{R_i}} \otimes \ket{j_i^{Q_i}}\bra{k_i^{Q_i}}) \Big)
\cr
&& =
\tr_{N+1}^Q{1\over D^{N+1}}\Big(I^{(N)} \otimes \Phi^{(N+1)}\Big)
    \Big( \sum_{{\bf j},{\bf k},k_{N+1}} \bigotimes_{i=1}^{N}
(\ket{j_i^{R_i}}\bra{k_i^{R_i}} \otimes \ket{j_i^{Q_i}}\bra{k_i^{Q_i}})
 \otimes \ket{k_{N+1}^{Q_{N+1}}}\bra{k_{N+1}^{Q_{N+1}}}  \Big)
\cr
&& =
{1\over D^{N+1}} \sum_{{\bf j},{\bf k},k_{N+1}} \Big(
\bigotimes_{i=1}^{N} (\ket{j_i^{R_i}}\bra{k_i^{R_i}}\Big) \otimes
\tr_{N+1}^Q\Phi^{(N+1)}\Big(
\bigotimes_{l=1}^{N}\ket{j_l^{Q_i}}\bra{k_l^{Q_i}}
 \otimes \ket{k_{N+1}^{Q_{N+1}}}\bra{k_{N+1}^{Q_{N+1}}}\Big)
\cr
&& =
{1\over D^{N+1}} \sum_{{\bf j},{\bf k},k_{N+1}} \Big(
\bigotimes_{i=1}^{N} (\ket{j_i^{R_i}}\bra{k_i^{R_i}}\Big) \otimes
\Phi^{(N)}\Big( \bigotimes_{l=1}^{N}\ket{j_l^{Q_i}}\bra{k_l^{Q_i}}
 \Big)
\cr
&& =
{1\over D^{N+1}}\Big(I^{(N)} \otimes \Phi^{(N)}\Big)
    \Big(  \sum_{{\bf j},{\bf k},k_{N+1}}
  \bigotimes_{i=1}^{N}
(\ket{j_i^{R_i}}\bra{k_i^{R_i}} \otimes \ket{j_i^{Q_i}}\bra{k_i^{Q_i}}) \Big)
\cr
&& =
{1\over D^N}\Big(I^{(N)} \otimes \Phi^{(N)}\Big)
    \Big(  \sum_{{\bf j},{\bf k}} \bigotimes_{i=1}^{N}
(\ket{j_i^{R_i}}\bra{k_i^{R_i}} \otimes \ket{j_i^{Q_i}}\bra{k_i^{Q_i}}) \Big)
\cr
&& =
J(\Phi^{(N)}) \;.
\end{eqnarray}

We have thus shown that $J(\Phi^{(N)})$, $N=1,2,\ldots$, form an exchangeable
sequence. According to the quantum de Finetti theorem for density operators
[see Eq.~(\ref{eq:definetti})], we can write
\begin{equation}   \label{eq:JdeFinetti}
J(\Phi^{(N)})  = \int d\rho\; p(\rho)\, \rho^{\otimes N} \;,
\end{equation}
where $p(\rho)\ge0$ is unique, and $\int d\rho\; p(\rho)=1$.
With the parameterization
\begin{equation}   \label{eq:rhoparam}
\rho = {1\over D} \sum_{l,j,m,k} S^{(1)}_{l,j,m,k} \ket{j^{R}}\bra{k^{R}}
   \otimes\ket{l^{Q}}\bra{m^{Q}} \;,
\end{equation}
Eq.~(\ref{eq:JdeFinetti}) takes the form
\begin{eqnarray}    \label{eq:Jparametrized}
J(\Phi^{(N)})
&=& {1\over D^N}\int_{{\cal D}} dS \; p(S) \;
\Big(\sum_{l,j,m,k} S^{(1)}_{l,j,m,k} \ket{j^{R}}\bra{k^{R}}
   \otimes\ket{l^{Q}}\bra{m^{Q}}  \Big)^{\otimes N} \cr
&=& {1\over D^N}\int_{{\cal D}} dS \; p(S) \;
\bigotimes_{i=1}^N \sum_{l_i,j_i,m_i,k_i} S^{(1)}_{l_i,j_i,m_i,k_i}
\ket{j_i^{R_i}}\bra{k_i^{R_i}}
   \otimes\ket{l_i^{Q_i}}\bra{m_i^{Q_i}} \cr
&=& {1\over D^N} \sum_{{\bf l},{\bf j},{\bf m},{\bf k}}
   \int_{{\cal D}} dS \; p(S) \;
\bigotimes_{i=1}^N S^{(1)}_{l_i,j_i,m_i,k_i}
  \ket{j_i^{R_i}}\bra{k_i^{R_i}}  \otimes\ket{l_i^{Q_i}}\bra{m_i^{Q_i}} \;,
\end{eqnarray}
where the integration variable is a vector with $D^4$ components,
$S=(S^{(1)}_{1,1,1,1},\ldots,S^{(1)}_{D,D,D,D})$, and where the integration
domain, ${\cal D}$, is the set of all $S$ that represent matrix
elements of a density operator. The function $p(S)$ is unique, $p(S)\ge0$,
and $\displaystyle{\int_{{\cal D}} dS\; p(S)=1}$.
Notice the slight abuse of notation in the first line of
Eq.~(\ref{eq:Jparametrized}), where the superscripts $R$ and $Q$ label the
entire sequences of systems $R_1,\ldots,R_N$ and $Q_1,\ldots,Q_N$,
respectively.

Comparing Eq.~(\ref{eq:Jparametrized}) with Eq.~(\ref{eq:Jdef}), we can
express the coefficients $S^{(N)}_{{\bf l},{\bf j},{\bf m},{\bf k}}$
specifying the quantum operation $\Phi^{(N)}$ [see Eq.~(\ref{eq:sljmk})]
in terms of the integral above:
\begin{equation}
S^{(N)}_{{\bf l},{\bf j},{\bf m},{\bf k}} = \int_{{\cal D}} dS \; p(S) \;
   \prod_{i=1}^N S^{(1)}_{l_i,j_i,m_i,k_i} \;.
\end{equation}
Hence, for any ${\bf j}$ and ${\bf k}$,
\begin{eqnarray}
\Phi^{(N)} \Big( \bigotimes_{i=1}^N \ket{j_i^{Q_i}}\bra{k_i^{Q_i}} \Big)
&=&  \sum_{{\bf l},{\bf m}} \int_{{\cal D}} dS \; p(S) \;
   \Big(\prod_{i=1}^N S^{(1)}_{l_i,j_i,m_i,k_i} \Big)
  \bigotimes_{i=1}^N \ket{l_i^{Q_i}}\bra{m_i^{Q_i}}    \cr
&=&  \int_{{\cal D}} dS \; p(S) \;
   \bigotimes_{i=1}^N
  \sum_{l_i,m_i} S^{(1)}_{l_i,j_i,m_i,k_i}\ket{l_i^{Q_i}}\bra{m_i^{Q_i}} \;.
\end{eqnarray}
The $D^4$ coefficients, $S^{(1)}_{l,j,m,k}$,  of the vector $S$ define
a single-system map, $\Phi_S$, via
\begin{equation}
\Phi_S(\ket{j^{Q}}\bra{k^{Q}})  \equiv
  \sum_{l,m} S^{(1)}_{l,j,m,k}\ket{l^{Q}}\bra{m^{Q}} \;\;\;(j,k=1,\ldots,D)\;.
\end{equation}
Hence
\begin{eqnarray}  \label{eq:predefinetti}
\Phi^{(N)} \Big( \bigotimes_{i=1}^N \ket{j_i^{Q_i}}\bra{k_i^{Q_i}} \Big)
&=& \int_{{\cal D}} dS \; p(S) \;
   \bigotimes_{i=1}^N
\Phi_S\Big(\ket{j_i^{Q_i}}\bra{k_i^{Q_i}}\Big) \cr
&=& \int_{{\cal D}} dS \; p(S) \;      \Phi_S^{\otimes N}
 \Big( \bigotimes_{i=1}^N \ket{j_i^{Q_i}}\bra{k_i^{Q_i}}\Big) \;.
\end{eqnarray}
Since this equality holds for arbitrary ${\bf j}$ and
${\bf k}$, it implies the representation
\begin{equation}  \label{eq:finaldefinetti}
\Phi^{(N)}  = \int_{{\cal D}} dS \; p(S) \;   \Phi_S^{\otimes N} \;.
\end{equation}
For all $S\in {\cal D}$, the map $\Phi_S$ is completely positive.
This can be seen by considering
$$J(\Phi_S) = (I\otimes\Phi_S)\big(\ket\Psi\bra\Psi\big) = {1\over D}
\sum_{l,j,m,k} S^{(1)}_{l,j,m,k} \ket{j^{R}}\bra{k^{R}}
\otimes\ket{l^{Q}}\bra{m^{Q}} \;,$$
which,  by definition of ${\cal D}$,  is a density operator and therefore
positive. It follows from a theorem by Choi
\cite{Choi1975} that $\Phi_S$ is completely positive.

To complete the proof, we will now show that $p(S)=0$ almost
everywhere (a.e.) unless $\Phi_S$ is trace-preserving, i.e., a
quantum operation.  More precisely, we show that if $U\in{\cal D}$ is
such that $\Phi_U$ is not trace-preserving, then there exists an open
ball $B$ containing $U$ such that $p(S)=0$ (a.e.) in $B\cap{\cal D}$.

For $\delta>0$ and $U\in{\cal D}$, we define $B_\delta(U)$ to be the
set of all $S$ such that $|S-U|<\delta$, i.e.,  $B_\delta(U)$ is the
open ball of radius $\delta$ centered at $U$. Furthermore, we define
$\bar B_\delta(U)=B_\delta(U)\cap{\cal D}$.

Let $U\in{\cal D}$ be such that $\Phi_U$
is not trace-preserving, i.e., there exists a density operator $\rho$ for
which $\tr[\Phi_U(\rho)]\ne1$. We distinguish two cases.

\vspace{3mm}\noindent{\bf Case (i)}:
$\tr[\Phi_U(\rho)]=1+\epsilon$, where $\epsilon>0$. Since $\tr[\Phi_S(\rho)]$
is a linear and therefore continuous function of the vector $S$, there exists
$\delta>0$ such that
\begin{equation}
\Big|\tr[\Phi_S(\rho)]-\tr[\Phi_U(\rho)]\Big|<\epsilon/2
\end{equation}
whenever $S\in B_\delta(U)$. For $S\in\bar B_\delta(U)$,
\begin{equation}
\tr[\Phi_S(\rho)]>1+\epsilon-\epsilon/2=1+\epsilon/2 \;.
\end{equation}
Therefore
\begin{eqnarray}
\tr\big[\Phi^{(N)}(\rho^{\otimes N})\big]
&=&\tr\left[\int_{{\cal D}} dS \; p(S) \; \Phi_S^{\otimes N}(\rho^{\otimes N})\right] \cr
&=&\int_{{\cal D}} dS \; p(S) \; \left(\tr[\Phi_S(\rho)]\right)^N \cr
&=&\int_{{\cal D}\backslash \bar B_\delta(U)}
          dS \; p(S) \; \left(\tr[\Phi_S(\rho)]\right)^N
   + \int_{\bar B_\delta(U)}  dS \; p(S) \; \left(\tr[\Phi_S(\rho)]\right)^N \cr
&\ge& \int_{\bar B_\delta(U)}  dS \; p(S) \; \left(\tr[\Phi_S(\rho)]\right)^N\cr
&>&   (1+\epsilon/2)^N \int_{\bar B_\delta(U)}  dS \; p(S) \;.
\end{eqnarray}
Unless $\displaystyle{\int_{\bar B_\delta(U)} dS \; p(S) =0}$, there exists $N$
such that $\tr\big[\Phi^{(N)}(\rho^{\otimes N})\big]>1$, which contradicts the
assumption that $\Phi^{(N)}$ is trace-preserving. Hence $p(S)=0$ (a.e.)
in $\bar B_\delta(U)$.

\vspace{3mm}\noindent{\bf Case (ii)}:
$\tr[\Phi_U(\rho)]=1-\epsilon$, where $0<\epsilon\le1$. Because of continuity,
there exists $\delta>0$ such that
\begin{equation}
\Big|\tr[\Phi_S(\rho)]-\tr[\Phi_U(\rho)]\Big|<\epsilon/2
\end{equation}
whenever $S\in B_\delta(U)$. Hence, for $S\in\bar B_\delta(U)$,
\begin{equation}
\tr[\Phi_S(\rho)]<1-\epsilon+\epsilon/2=1-\epsilon/2 \;.
\end{equation}
Now assume that $\displaystyle{\int_{\bar B_\delta(U)} dS \; p(S) =\eta>0}$.
Then, letting $N=1$,
\begin{eqnarray}
1 = \tr\big[\Phi^{(1)}(\rho)\big]
&=&\tr\left[\int_{{\cal D}} dS \; p(S) \; \Phi_S(\rho)\right] \cr
&=&\int_{{\cal D}\backslash \bar B_\delta(U)}
          dS \; p(S) \; \tr[\Phi_S(\rho)]
   + \int_{\bar B_\delta(U)}  dS \; p(S) \; \tr[\Phi_S(\rho)] \cr
&<&
\int_{{\cal D}\backslash \bar B_\delta(U)}
    dS \; p(S) \; \tr[\Phi_S(\rho)] + \eta(1-\epsilon/2) \;,
\end{eqnarray}
which implies that
\begin{equation}
\int_{{\cal D}\backslash \bar B_\delta(U)} dS\;p(S)\;\tr[\Phi_S(\rho)] > 1-\eta +
\eta\epsilon/2 > 1-\eta \;.
\end{equation}
Since
\begin{equation}
\int_{{\cal D}\backslash \bar B_\delta(U)} dS\;p(S)=1-\eta \;,
\end{equation}
it follows that there exist $\zeta>0$ and a point $V\in {\cal D}\backslash \bar
B_\delta(U)$ such that $\tr[\Phi_V(\rho)]>1$ and
\begin{equation}
\int_{\bar B_\xi(V)} dS\;p(S) > 0 \;\;\mbox{ for all } \xi\le\zeta
\;.
\end{equation}
We are thus back to case (i) above. Repeating the argument of case
(i) one can show that this contradicts the assumption that
$\Phi^{(N)}$ is trace preserving for large $N$. It follows that
$\eta=0$, i.e., $p(S)=0$ (a.e.) in $\bar B_\delta(U)$. This concludes
the proof of the de Finetti theorem for quantum operations.

\section{Concluding Remarks}

What we have proven here is a representation theorem.  It shows us
when an experimenter is warranted to think of his (prior) {\it
known\/} quantum operation assignment as built out of a lack of
knowledge of a ``true'' but {\it unknown\/} one.  In that way, the
theorem has the same kind of attraction as the previous quantum de
Finetti theorem for quantum states
\cite{Hudson1976,Caves2002b,Emch2002}.

In particular for an information-based interpretation of quantum
mechanics such as the one being developed in
Refs.~\cite{Caves2002a,Caves2002b,Fuchs2002}, it may be a necessary
ingredient for its very consistency.  In
Refs.~\cite{Fuchs2002,FuchsWHAT}, it has been argued strenuously that
quantum operations should be considered of essentially the same
physical meaning and status as quantum states themselves:  They are
Bayesian expressions of an experimenter's judgment.  This could be
captured in the slogan ``a quantum operation is really a quantum
state in disguise.'' In other words, the Choi representation theorem
\cite{Choi1975} is not just a mathematical nicety, but is instead of
deep physical significance.

Therefore, just as an unknown quantum state is an oxymoron in an
information-based interpretation of quantum mechanics, so should be
an unknown quantum operation.  In the case of quantum states, the
conundrum is solved by the existence of a de Finetti theorem for
quantum tomography.  Here we have shown that the conundrum in quantum
process tomography can be solved in almost the same way.  One might
reject the arguments leading to the slogan that a quantum operation
is a quantum state in disguise \cite{CavesMerminCorrespondence}, but
then one should be curious about the nice fit of the formalism to the
philosophy.

\acknowledgments

CAF and RS acknowledge the hospitality of the Australian Special
Research Centre for Quantum Computer Technology at the University of
Queensland, where part of this work was carried out Spring 2002. PFS
acknowledges support from DIMACS, Rutgers University and thanks Bell
Labs, Lucent Technologies for its warm hospitality.  CAF also thanks
G. L. Comer for enlightening discussions on the day of this paper's
completion.


\end{document}